\documentclass{article}

\PassOptionsToPackage{numbers, compress}{natbib}

\usepackage[preprint]{neurips_2024}

\usepackage[utf8]{inputenc} 
\usepackage[T1]{fontenc}    
\usepackage{hyperref}       
\usepackage{url}            
\usepackage{booktabs}       
\usepackage{amsfonts}       
\usepackage{nicefrac}       
\usepackage{microtype}      
\usepackage{xcolor}         
\usepackage{multirow}
\usepackage{booktabs}
\usepackage{tabularx}
\usepackage{diagbox}

\usepackage{graphicx}
\usepackage{bbding}

\usepackage{float}
\usepackage{graphicx}
\usepackage{amsmath, amssymb, amsfonts}
\usepackage{ifsym}

\newcommand{\E}{\mathbb{E}}

\title{Identity Preserving Latent Diffusion for Brain Aging Modeling}
\author{
	Gexin Huang\thanks{Co-author.} \\
	University of British Columbia \\
	\texttt{gexinml@gmail.com} \\
	\And
	Zhangsihao Yang\footnotemark[1] \\
	Arizona State University \\
	\texttt{zshyang1106@gmail.com} \\
	\And
	Yalin Wang \\
	Arizona State University \\
	\texttt{ylwang@asu.edu} \\
	\AND
	Guido Gerig \\
	University of New York \\
	\texttt{gerig@nyu.edu} \\
	\And
	Mengwei Ren \\
	University of New York \\
	\texttt{mengwei.ren@nyu.edc} \\
	\And
	Xiaoxiao Li\thanks{Corresponding author.} \\
	University of British Columbia \\
	\texttt{xiaoxiao.li@ece.ubc.ca} \\
}

\begin{document}

	\maketitle

	\begin{abstract}
		Structural and appearance changes in brain imaging over time are crucial indicators of neurodevelopment and neurodegeneration. The rapid advancement of large-scale generative models provides a promising backbone for modeling these complex global and local changes in brain images, such as transforming the age of a source image to a target age. However, current generative models, typically trained on independently and identically distributed (i.i.d.) data, may struggle to maintain intra-subject spatiotemporal consistency during transformations.
		We propose the Identity-Preserving Longitudinal Diffusion Model (IP-LDM), designed to accurately transform brain ages while preserving subject identity. Our approach involves first extracting the identity representation from the source image. Then, conditioned on the target age, the latent diffusion model learns to generate the age-transformed target image. To ensure consistency within the same subject over time, we regularize the identity representation using a triplet contrastive formulation.
		Our experiments on both elderly and infant brain datasets demonstrate that our model outperforms existing conditional generative models, producing realistic age transformations while preserving intra-subject identity.

	\end{abstract}

	\section{Introduction}
	
	Modeling brain aging is crucial for understanding neurological conditions and the overall impact of aging on brain structure and function.
	This knowledge is essential for early diagnosis, monitoring disease progression, and developing effective treatments. 
	Generative models have emerged as a promising tool for simulating the complex process of brain aging, offering the potential to generate realistic age-progressed images. However, existing generative models encounter significant challenges for modeling brain development. 
	Current image-to-image generative models assume independent and identically distributed (i.i.d.) data, fall short as they do not consider the non-i.i.d. properties inherent in longitudinal datasets, where images of the same subject are collected over time.
	
	While longitudinal representation learning has been extensively studied~\cite{ren2022local,chaitanya2020contrastive,ouyang2021self,to2021self,wei2021consistent} to incorporate spatiotemporal consistency in the latent space, it has predominantly been applied to global and local downstream tasks such as classification and segmentation~\cite{couronne2021longitudinal,zhao2021longitudinal}, and has not been extensively integrated into pixel/voxel level image synthesis task for individual progression, leaving a gap between spatiotemporal consistent representation and the synthesis power of generative models. 
	Additionally, video generative models~\cite{blattmann2023stable,brooks2022generating,blattmann2023align,gu2023reuse,guo2023animatediff,he2022latent,wang2023modelscope} that account for non-i.i.d. data typically deal with densely sampled frames, making them unsuitable for longitudinal brain imaging datasets, which are characteristically sparse with fewer time points.
	
	To overcome these limitations, we propose a novel generative model specifically designed to simulate brain development while preserving intra-subject identity. Our approach builds on a latent diffusion model conditioned on both age and subject identity, ensuring that the generated images reflect the aging process of individual subjects. The identity representation within our model is regularized through a triplet contrastive representation learning formulation, which enhances the model's ability to maintain consistent identity features across different ages. Our contributions are three-fold: (1) We propose an age- and identity-conditioned latent diffusion model that transforms the appearance of a single input brain image to reflect arbitrary and continuous age changes; (2) We incorporate triplet contrastive constraints to ensure consistent intra-subject identity representation; (3) Our results on both elderly and infant brain datasets demonstrate the effectiveness of our method in synthesizing high-quality brain aging transformations while preserving subject identity.

	
	\section{Related Work}
	
	%
	
	\noindent\textbf{Brain Aging Modeling.}
	Understanding brain aging is crucial for studying neurodevelopmental and neurodegenerative diseases and developing effective interventions. Traditional brain aging models often rely on linear and non-linear regression methods to predict age-related changes in brain morphology and function~\cite{franke2010estimating,sivera2019model,long2012healthy,blinkouskaya2021brain,huizinga2018spatio}. These methods are typically task-specific, developed to track particular regions of interest, such as brain tumor growth~\cite{domschke2014mathematical,elazab2014content,swan2018patient,yuan2016brain}. 
	With the advent of deep learning, end-to-end models have been developed for broader age-related global tasks, such as age~\cite{sihag2024explainable,mouches2021unifying,zhao2019variational,ijishakin2024semi} and disease~\cite{lee2022deep,khan2021machine,bashyam2020mri,yin2023anatomically} prediction. 
	Specifically, these studies~\cite{mouches2021unifying,zhao2019variational,ijishakin2024semi} also demonstrate their models' capabilities in longitudinal brain image generation.
	\noindent\textbf{Image-to-Image Models} are trained to transform an input source image towards a task-specific target image, e.g., style transfer~\cite{meng2021sdedit,zhang2023inversion}, local/global image editing~\cite{avrahami2022blended,avrahami2023blended}, image enhancement. 
	With the advancement of generative models, image-to-image task can be formulated as an image-conditioned generative process, where the backbone model (GAN~\cite{pix2pix2017,liu2017unsupervised,yi2017dualgan,zhu2017unpaired,huang2018multimodal,xiong2019consistent,emami2020spa}, diffusion model~\cite{ramesh2022hierarchical,rombach2022high,saharia2022photorealistic,nichol2021glide,brooks2023instructpix2pix}) conditionally takes an input image as the generation prior. These models are trained with pairwise datasets under a supervised formulation. However, most existing methods are trained with i.i.d. data, where the intra-subject correlation may not be properly modeled and preserved. Thus, the extension to longitudinal brain images remains nontrivial.

	\noindent\textbf{Image-to-Image Models for Aging} have recently been empowered by the use of data-driven generative models. 
	Earlier works re-purpose conditional GANs~\cite{yang2018learning,wang2018face,he2019s2gan,or2020lifespan} for face-aging prediction, by disentangling the representation of age and identity. 
	Similarly, ~\cite{xia2021learning} explores the application of conditional GANs for aging brain synthesis, without relying on longitudinal data. 
	~\cite{peng2021longitudinal} further utilize longitudinal MRI datasets to predict longitudinal infant MR images at a predefined age group for data imputation.
	Recently,~\cite{pinaya2022brain} investigates the use of Latent Diffusion Models~\cite{rombach2022high} for covariate-to-image synthesis where
	cross-sectional T1-weighted MRI images are sampled given imaging covariates including age. However, its further extension on image-conditioned synthesis remain unexplored.  SADM~\cite{yoon2023sadm} develop a pixel-level diffusion model along with a sequence transformer, to predict the brain aging in an autoregressive manner. The image for each target age group is predicted from its preceding group, which may be limited in arbitrary age transformation, whereas our method enables continuous age transformation without sequential modeling.

	\section{Method}
\begin{figure*}[htbp]
		\centering
		\includegraphics[width=1\linewidth]{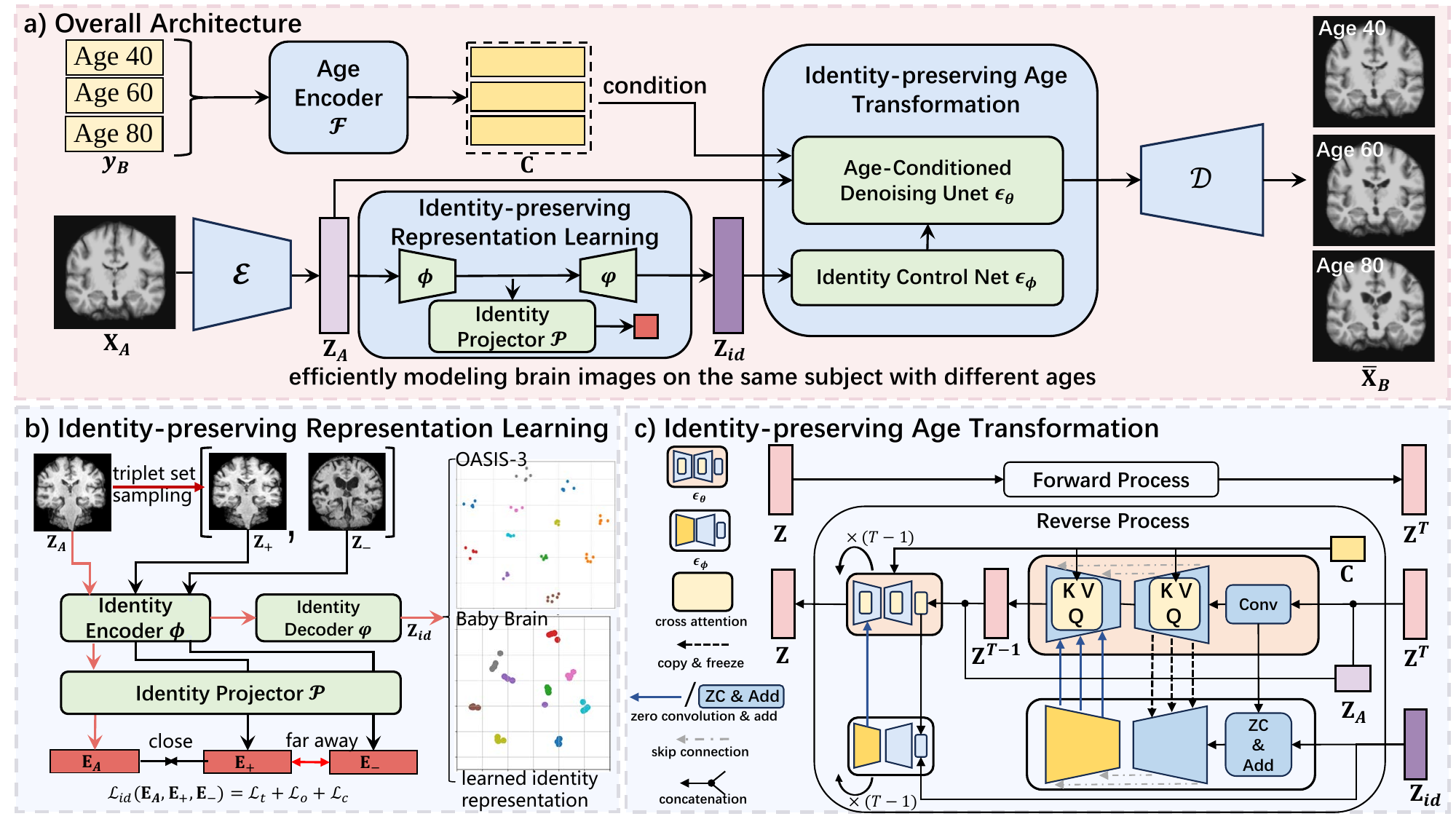}
		\caption{Overview of proposed longitudinal diffusion model. a) shows the overall architecture of IP-LDM, consisting of an age and identity conditioned latent diffusion, wherein $\mathcal{E}$ and $\mathcal{D}$ are the image encoder and decoder, respectively. b) illustrates the details of the identity representation learning along with the learned feature distributions (different color indicates different subjects). c) depicts the forward and backward process in the latent manipulation module.}
		
		\label{fig_2} 
	\end{figure*}

	Fig. \ref{fig_2} gives an overview of our framework (red outline) alongside details of tailored modules in the framework (blue outline). 
	Once trained, out model enables the prediction of intra-subject brain images under arbitrary age, by taking a source image along with a target age as the input.

	\noindent\textbf{Overview.}
	Given a set of structural magnetic resonance imaging (sMRI) images $\mathbf{X}$, the generation of longitudinal brain images \cite{oh2022learn} involves taking a source brain image $\mathbf{X}_{A}$ and generating a target image $\mathbf{X}_{B}$ based on the target age $\mathbf{y}_{B}$, as shown in Fig.~\ref{fig_2}. Our IP-LDM initially uses a visual encoder $\mathcal{E}$, trained from a pixel-space autoencoder, to extract semantic information and compress $\mathbf{X}_{A}$ into latent semantic features $\mathbf{Z}_{A}$. Next, an identity-preserving representation learning (IRL, Fig.~\ref{fig_2} (b)) is designed to extract the identity feature $\mathbf{Z}_{id}$. The IRL comprises an identity auto-encoder (comprising an encoder $\phi$ and a decoder $\varphi$) and an identity projector, which regularizes the latent features by ensuring identity preservation to yield the identity feature $\mathbf{Z}_{id}$. Concurrently, an age encoder computes the age embedding $\mathbf{C}$ from $y_{B}$. Finally, an identity-preserving age transformation (IAT, Fig.~\ref{fig_2} (c)) operates under the latent diffusion model framework, which includes an age-conditioned denoising U-Net and an identity control net. The age-conditioned denoising U-Net integrates $\mathbf{Z}_{A}$ into Gaussian noise and uses $\mathbf{C}$ as a condition to progressively recover the target latent semantic features, incorporating $\mathbf{Z}_{id}$ via the identity control net. The visual decoder $\mathcal{D}$ from the autoencoder then reconstructs the target brain image $\bar{\mathbf{X}}_{B}$ with the specified age while preserving identity consistency in brain structures.

	\vspace{0.1cm}

	\subsection{Preliminaries on Latent diffusion model}
	Diffusion models (DMs) are probabilistic generative models that comprise two essential processes: the forward process (known as the diffusion process) and the reverse process, which restore a sampled variable from Gaussian noise to a sample of the learned data distribution via iterative denoising. Given training data, the \textit{forward process} destroys the structure of the data by gradually adding Gaussian noise. The sample at each time point is defined as $\mathbf{X}_t=\sqrt{\alpha_t}\mathbf{X}_0 + \sqrt{1- \alpha_t}\epsilon$ where $\mathbf{X}_t$ is a noisy version of input $\mathbf{X}_0$, $t \in \{1, \cdots, T \}$, $\alpha$ is a hyperparameter to control the the variance of the additive pre-scheduled noise, and $\epsilon \sim \mathcal{N}(0, \mathbf{I})$. The \textit{reverse process} is modeled by applying a neural network $\epsilon_{\theta}(\mathbf{x}_t, t)$ to the samples at each step to recover the original input. The learning objective is $ \epsilon_{\theta}(x,t) \approx \epsilon_{t}$ ~\cite{ho2020denoising}, in which neural networks $\epsilon_{\theta}$ is commonly built by the U-Net.
	
	Latent diffusion models (LDMs) compress the input using an autoencoder (AE), which overcomes the computationally expensive limitation of DMs due to the operation in the pixel space. Specifically, the AE is first trained with the brain images to compress the high-resolution images into a lower-dimensional latent representation. The DM is sequentially trained to generate its latent representation $\mathbf{Z}$ using the U-Net. Additionally, the LDM can be generalized to a \textit{conditional} one by inserting auxiliary input into the neural network $\epsilon_{\theta}$. When we start from the source image with input conditions, we can generate new age conditional images by editing the image. In this image-to-image translation, the degree of degradation from the original image is controlled by a parameter that can be adjusted to preserve either the semantic content or the appearance of the original image.
	
	IP-LDM is based on the conditional latent diffusion model, composed of four main components: an AE, a condition encoder (i.e., the age encoder), an identity preservation module, and a DM-based age transformation module. 
	Based on previous work~\cite{yu2021vector}, IP-LDM adopts the AE that comprises the visual encoder $\mathcal{E}$ and decoder $\mathcal{D}$ as shown in Fig.~\ref{fig_2}, which is trained with a combination of L1 loss, perceptual loss~\cite{zhang2018unreasonable}, a patch-based adversarial objective~\cite{isola2017image}. More precisely, given a brain image $\mathbf{X}\in \mathcal{R}^{H\times W \times 1}$ in the grey-scale space, the encoder $\mathcal{E}$ encodes $\mathbf{X}$ into latent semantic features $\mathbf{Z}= \mathcal{E}(\mathbf{X})$. Then, the decoder $\mathcal{D}$ reconstructs the image from the latent, giving $\bar{\mathbf{X}}=\mathcal{D}(\mathbf{Z}) = \mathcal{D}(\mathcal{E}(\mathbf{X}))$, where
	$\mathbf{Z} \in \mathcal{R}^{h\times w \times c}$. Notice that the encoder downsamples the image to a two-dimensional structure features with a factor $f = H/h = W=w$, effectively preserving the inherent spatial structure of $\mathbf{X}$ to achieve a better semantic representation extraction.
	
	\subsection{Age-conditioned LDM}
	
	To manipulate the source brain image to match the target brain image according to a specific age, we first incorporate age information as a condition to construct the conditional Latent Diffusion Model (LDM).
	The age encoder, denoted as $\mathcal{F}$, is designed to encode the continuous age condition into an embedding vector, enabling precise control over the brain age of generated images. Specifically, $\mathcal{F}$ encodes the age condition from a continuous space onto a manifold, where brain age is represented as a continuous variable spanning the entire lifespan (e.g., from 1.8 months to 91 years). Compared to building category embeddings that divide lifespan age into several clusters, encoding this continuous representation facilitates the generation of realistic and age-appropriate brain images. This is particularly beneficial for studying developmental changes and age-related brain alterations, as it allows for a more nuanced and accurate portrayal of brain aging. This encoder is based on the work~\cite{dey2021generative}, leveraging a four-layer Multi-Layer Perceptron (MLP) with ReLU activation functions and instance adaptive normalization.
	Initially, the age encoder normalizes the age condition to the range [0, 1]. Subsequently, it outputs the age embeddings $\mathbf{C} = \mathcal{F}(y_{B}) \in \mathbb{R}^{d}$, where $d$ represents the dimensionality of the age embeddings, which matches the dimensionality of the hidden features in the U-Net. These embeddings $\mathbf{C}$ are then used as conditional inputs to the age transformation model, thereby guiding the image generation process within the diffusion model.

	\subsection{Identity-preserving Representation Learning}
	IP-LDM aims to preserve the identity information of brains during the generation of brain age transformation. However, the standard diffusion model only generates brains that follow the target data distribution, consequently, the generated brains may match any subject in the target age group. In other words, using the diffusion model alone can not guarantee that the generated brains can preserve the identity information. Therefore, IRL is designed to extract the identity features from the source brain images, which are sequentially incorporated into the diffusion model to maintain the identity information of the brain on the target age transformation. 
	The training process for the IRL is illustrated in Fig. \ref{fig_2}(b).
	
	The triplet set sampling strategy is first employed to train the IRL for effective identity preservation. Specifically, the source image \(\mathbf{X}_{A}\) is selected as the anchor while a positive sample \(\mathbf{X}_{+}\) is randomly selected from images of the same identity. In contrast, a negative sample \(\mathbf{X}_{-}\) is randomly sampled from images of different identities. Considering the lifespan brain aging dataset is a long-tail distribution, i.e., the number of images with younger and older ages is smaller than the number of images with normal ages, we additionally adopt the weighted random sampling to re-balance the constructed triplet set. This triplet set sampling ensures that the IRL learns to distinguish between images of the same identity and those of different identities.
	
	Next, the triplet set is compressed to the latent space via $\mathcal{E}$. The IRL designs an identity encoder $\phi$ and identity decoder $\varphi$ to compose a bottleneck network,  which is built using stacked 2-dimensional convolution layers and deconvolution layers, respectively. This aims to extract more compact and discriminative representations from the latent space, formulated as $\phi(\mathcal{E}(\mathbf{Z}))$. Sequentially, an identity projector, consisting of a stack of fully connected layers with the ReLU activation function, is constructed to obtain triplet identity embeddings {$\mathbf{E}_{A}$, $\mathbf{E}_{+}$, $\mathbf{E}_{-}$}, whose architecture is based on~\cite{chen2020simple} for better identity preservation learning. Finally, the triplet identity embeddings are constrained to minimize the following loss functions.
	
	\textbf{Triplet loss.} The loss encourages the embedding vectors of the anchor and positive images to be close to each other, while simultaneously pushing the embedding vector of the negative image further away. It is formulated as:
	\begin{equation}
	\mathcal{L}_{t}(\mathbf{E}_{A}, \mathbf{E}_{+}, \mathbf{E}_{-}) = \max \left( \| \mathbf{E}_{A} - \mathbf{E}_{+} \|^2_{F} - \| \mathbf{E}_{A} - \mathbf{E}_{-} \|^2_{F} + \alpha, 0 \right),
	\end{equation}
	wherein the $\alpha$ is the margin that specifies the minimum desired distance between the positive and negative pairs. During training, the model iteratively adjusts the identity embedding space so that the distance between the anchor and positive images is minimized, while the distance between the anchor and negative images is maximized. This process helps the model learn to discriminate between different identities based on their unique features.
	
	\textbf{Cosine Similarity Loss.} This loss enhances the IRL's ability to learn the identity similarity between the anchor and the positive sample, formulated as
	\begin{equation}
	\mathcal{L}_{o}(\mathbf{E}_{A}, \mathbf{E}_{+}) = 1 - \frac{{\mathbf{E}_{A} \cdot \mathbf{E}_{+}}}{{\|\mathbf{E}_{A}\|_2 \|\mathbf{E}_{+}\|_2}}.
	\end{equation}
	
	\textbf{Collapse Regularization.} The regularization ensures that the individual dimensions of the feature from the same identity are uncorrelated to avoid collapsed solutions, i.e., all outputs of the IRL are equal. It is formulated as: 
	\begin{equation}
	\mathcal{L}_{c}(\mathbf{E}_{A}, \mathbf{E}_{+}) = \gamma \| \mathbf{E}_{A}^\top\mathbf{E}_{+} - \mathbf{I}\|^2_{F},
	\end{equation}
	wherein $\gamma$ is the regularization weight and $\mathbf{I}$ is the unit matrix. Sequentially, the identity loss is formulated as $\mathcal{L}_{id}(\mathbf{E}_{A}, \mathbf{E}_{+}, \mathbf{E}_{-})=\mathcal{L}_{t}(\mathbf{E}_{A}, \mathbf{E}_{+}, \mathbf{E}_{-})+\mathcal{L}_{o}(\mathbf{E}_{A}, \mathbf{E}_{+})+\mathcal{L}_{c}(\mathbf{E}_{A}, \mathbf{E}_{+})$. As a result, the output of the IRL contains information about the local structures and the general shape of the brain, which play a key role in generating the same identity.

	\subsection{Identity-preserving Age Transformation}
	IP-LDM aims to generate the target brain images in the fashion of the conditional latent diffusion model, which is able to efficiently generate high-fidelity images without facing the mode collapse that usually occurs in the generative adversarial network (GAN).
	Thus, the IAT is designed to generate the brain age transformation, which is composed of the conditioned denoising UNet $\epsilon_{\theta}$ and the identity control net $\epsilon_{\phi}$. Specifically, the conditioned denoising UNet $\epsilon_{\theta}$ is designed to integrate semantic features of the source image $\mathbf{Z}_{A}$ and the age embeddings $\mathbf{C}$ to generate the target brain image according to the target age. The identity control net $\epsilon_{\phi}$ is designed to leverage the identity features $\mathbf{Z}_{id}$ to assist the conditioned denoising UNet in maintaining identity consistency over the age transformation during the reverse process. 
	
	As shown in Fig.~\ref{fig_2} (c), IAT destroys the structure of source latent features $\mathbf{Z}_{A}$ via the forward process during training. After that, IAT will leverage the $\epsilon_{\theta}$ to iteratively generate the target latent features $\mathbf{Z}_{b}$ over $T$ time steps via the reverse process. Based on the previous work~\cite{rombach2022high}, the age condition is incorporated with the cross-attention mechanism in $\epsilon_{\theta}$. Specifically, the age embedding $\mathbf{C}$ is mapped to the intermediate layers of the UNet via a cross-attention layer implementing $ Attention(Q; K; V ) = softmax(\frac{QK^\top}{\sqrt{d}}) \cdot V$, with $Q=\mathbf{W}_{Q}^{(i)} \cdot \psi_(\mathbf{Z}_t), K=\mathbf{W}_{K}^{(i)}\cdot \mathbf{C}, V=\mathbf{W}_{V}^{(i)}\cdot \mathbf{C}$. Note that, $\psi_(\mathbf{Z}_t)$ is a (flattened) intermediate representation of the U-Net $\epsilon_{\theta}$ and $\mathbf{W}_{Q}$, $\mathbf{W}_{K}$, and $\mathbf{W}_{V}$ are learnable projection matrices in the $i$-th layer.
	
	We first train $\epsilon_{\theta}$ to enhance the generation capability and model stabilization, which is under the objective function:
	\begin{equation}
	\mathcal{L}_{}==\E_{\mathbf{Z}_A, y_{A}, t,\epsilon\sim \mathcal{N}(0,1)} \left [||\epsilon-\epsilon_{\theta}(\mathbf{Z}_t, y_{A}, t)||^2\right],
	\end{equation}
	wherein $\mathbf{Z}_t$ is the denoised $\mathbf{Z}_{A}$ in $t$-th step.
	
	After that, instead of generating the target latent features from the Gaussian noise, IAT is designed to concatenate the source latent features into the noised features as the input of $\epsilon_{\theta}$, which is capable of fully leveraging the semantic information of source images. Then, a convolution neural network is leveraged to integrate the two features into the new ones, wherein the input of $\epsilon_{\theta}$ in $t$-th reverse process step is formulated as $\tilde{\mathbf{Z}}_t = Conv([\mathbf{Z}_t, \mathbf{Z}_A])$. 
	
	Furthermore, to precisely control the identity information of brains, IAT adopts the identity control net $\epsilon_{\phi}$, based on \cite{zhang2023adding}, to insert the identity features into the decoder of $\epsilon_{\theta}$. Specifically, $\epsilon_{\phi}$ utilizes a trainable copy of the original weights of the pre-trained $\epsilon_{\theta}$. The trainable copy and the original frozen model are connected with the zero l convolution layer $\mathcal{Z}(\cdot)$, where the weights are initialized as zeros and no noise is added in the learning process. 
	$\epsilon_{\phi}$ applies the copy to each encoder level of the U-net $\epsilon_{\theta}$ and incorporates the zero convolution layer to yield the outputs, which are integrated into the decoder layer of the U-net $\epsilon_{\theta}$. The integrated feature in $l$-th layer is formulated as:
	\begin{equation}
	z_{id}^{l+1}=\epsilon_{\theta}(z^{l}) + \mathcal{Z}_{z2}(\epsilon_{\phi}(z^{l}+\mathcal{Z}_{z1}(z_{id}^{l}))),     
	\end{equation}
	where $z_{id}^{l}=\mathbf{Z}_{id}$, $z^{l}$ is the feature maps of the $l$-th encoder layer in $\epsilon_{\theta}$, and the two zero convolution layers are built by two parameters z1 and z2 respectively. As a result, we jointly train pre-trained $\epsilon_{\theta} $ and $\epsilon_{\phi}$ under the optimization objective is formulated as:
	\begin{equation}
	\mathcal{L}=\E_{\mathbf{Z}_A, y_{B}, Z_{id}, t,\epsilon\sim \mathcal{N}(0,1)} \left [||\epsilon - \epsilon_\theta([\tilde{\mathbf{Z}}_t, y_{B}, Z_{id}, t)||^2\right].
	\end{equation}
	



	\section{Experiments}

	\subsection{Datasets, implementation details, and evaluation.}

	\textbf{Data and Preprocessing.} \underline{OASIS-3} \cite{lamontagne2019oasis} is a publicly available dataset comprising 1639 brain MRI scans of 992 subjects, each with 1–5 temporal acquisitions over a 5-year observation window. The cohort includes individuals aged 42 to 97, featuring both cognitively normal and mildly impaired individuals, as well as those with Alzheimer’s Disease. \underline{Baby Brain} \cite{ren2022local} is an infant brain imaging study that longitudinally acquires 1272 structural T1w/T2w MRIs from 552 infants, including controls and high-risk infants for Autism Spectrum Disorder (ASD) \cite{lord2018autism}, over the age range of 3 to 36 months. We preprocess the OASIS-3 dataset by cropping to a size of [160, 160, 198] to remove redundant background and selecting the middle slice along the last axis to form images with a resolution of $160 \times 160$. These images are normalized to a [0, 1] range without registration to preserve age-related shape deformation. Baby Brain is cropped to [160, 196, 128], with the middle slice along the second axis resized to [200, 160]. Other preprocessing follows the OASIS-3 protocol.
	
	\textbf{Implementation.} We employ the Adam optimizer with $\beta_1 = 0.9$ and $\beta_2 = 0.999$. For Baby Brain, we train the autoencoder (AE) with a batch size of 256 for 20,000 steps and a learning rate of $1 \times 10^{-4}$. For OASIS-3, the AE is trained with a batch size of 320 for 20,000 steps at the same learning rate. 
	In our experiments, we found that Larger batch sizes significantly improve outcomes, and select the AE to prevent image blur induced by the Kullback-Leibler (KL) divergence regularization.
	Both datasets' U-Nets are trained with a batch size of 256 for 20,000 steps, using a learning rate of $1 \times 10^{-4}$. For the manipulation module, the batch size is reduced to 128 for 10,000 steps with a learning rate of $1 \times 10^{-5}$, due to the pre-trained U-Net. All experiments are conducted on a single A100 GPU.
	
	\textbf{Evaluation.} We compare our model with three baselines—cGAN~\cite{ren2021q}, DAE~\cite{preechakul2022diffusion}, and InstructPix2Pix~\cite{brooks2023instructpix2pix}—covering both GAN-based (cGAN) and diffusion-based (DAE, InstructPix2Pix) methods. 
	Evaluation is performed on the test dataset, using MRI images taken at different ages as target images. Specifically, for each subject, we consider pairs of MRI images from different ages (e.g., 2 months and 10 months) and test image generation in both translation directions.
	We evaluate the generated images using the following metrics: Structural Similarity Index (SSIM), Peak Signal-to-Noise Ratio (PSNR), Fréchet Inception Distance (FID), Kernel Inception Distance (KID), Root Mean Square Error (RSMSE), and Adjusted Rand Index (ARI). Details on the metric computations are provided in App.~\ref{supp_sec:metrics}.

	\subsection{Benchmark Performance} 

	\begin{figure*}[h]
		\centering
		\includegraphics[width=\linewidth]{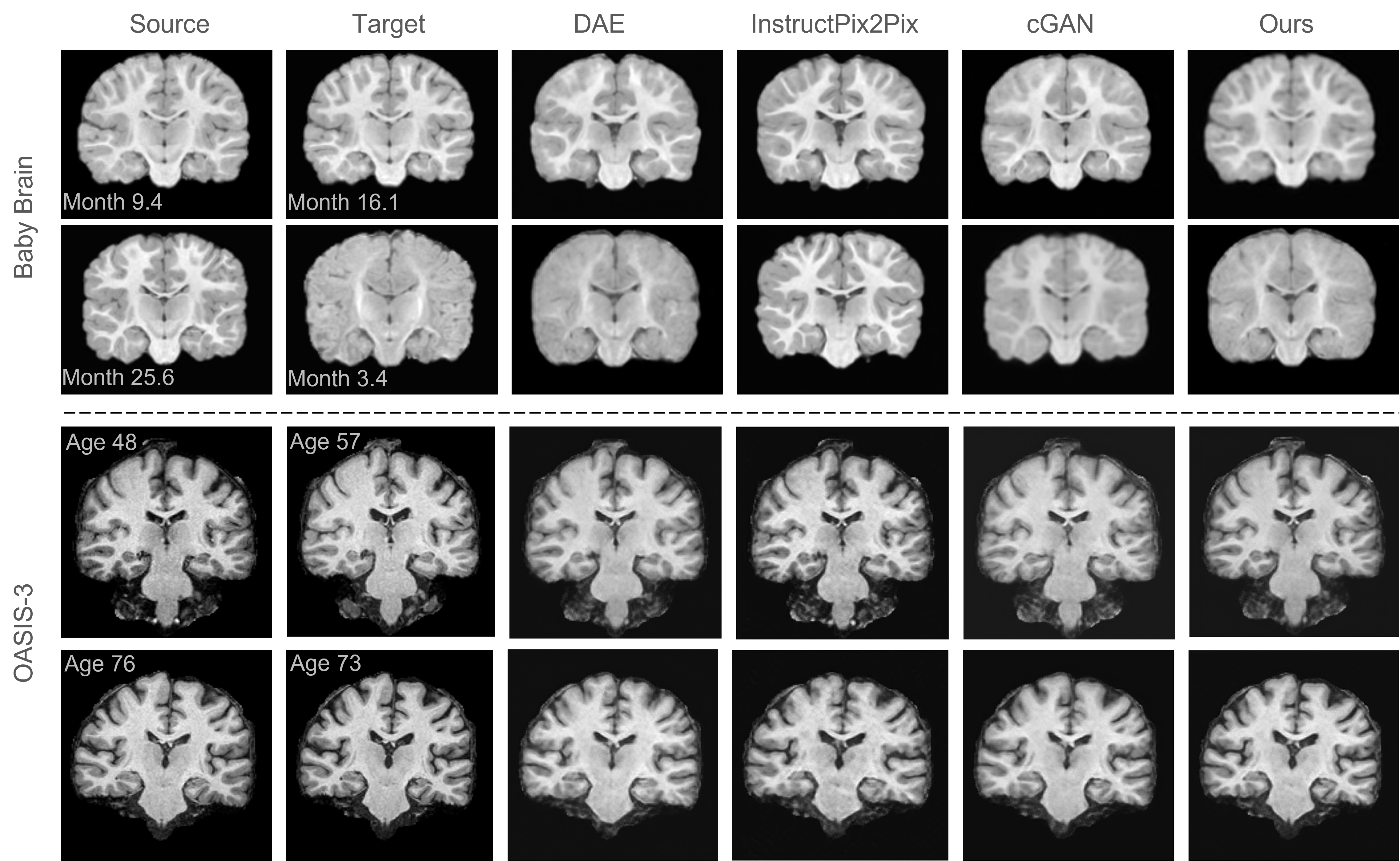}
		\vspace{-10pt}
		\caption{Visualization of baseline comparison on the Baby Brain and OASIS-3 datasets. The brain images are generated by different models based on the source images, which are expected to align with the target images.
			In the first row, the source and target are similar in age, resulting in subtle yet discernible longitudinal changes. 
			Our method preserves the identity more effectively compared to other non-identity-preserving baselines.
		}
		\label{fig:baseline}
	\end{figure*}
	
	

	
	Tab.~\ref{table1} presents a performance comparison of IP-LDM against SOTA methods on the OASIS-3 and Baby Brain datasets. On the OASIS-3 dataset, IP-LDM achieves the highest SSIM 0.949 and PSNR 35.15, indicating superior structural similarity and image fidelity. Additionally, IP-LDM records the lowest FID 4.733 and RMSE 1.868, reflecting high-quality and accurate image generation. The highest ARI 0.99 further underscores IP-LDM's capability in maintaining identity preservation. Similarly, on the IBIS dataset, IP-LDM outperforms other methods with the highest SSIM 0.674 and PSNR 32.989, and the lowest FID 4.984 and RMSE 8.996, demonstrating its robustness in producing realistic and precise brain images during age transformation. Notably, the GAN-based method (cGAN) performs better on the Baby Brain dataset compared to the OASIS-3 dataset. 
	We attribute this difference to the inherent characteristics of the datasets. Please refer to App.~\ref{supp_sec:age_dist} for further details.
	
	The visualization shown in Fig.~\ref{fig:baseline} further underscores the superior performance of IP-LDM. The figure displays brain images generated by different methods across the Baby Brain and OASIS-3 datasets, generating target images based on source images. For both datasets, IP-LDM produces images that closely resemble the target images, maintaining fine structural details and anatomical accuracy. The generated images exhibit clear ventricles and well-preserved brain structures, indicative of successful identity preservation and realistic aging transformation. In contrast, images generated by cGAN and DAE display noticeable artifacts and structural inconsistencies. cGAN, while producing visually realistic images, often fails to maintain finer identity-specific details, leading to less accurate age transformations. DAE struggles with both realism and structural integrity, showing blurred and less detailed images. InstructPix2Pix performs better than cGAN and DAE but still falls short of the accuracy and fidelity demonstrated by IP-LDM, especially in maintaining subtle geometric variations and anatomical features.
	
	\subsection{Quantitative and qualitative identity preservation through age transformation.}
	
\begin{figure*}[h]
		\centering
		\includegraphics[width=\linewidth]{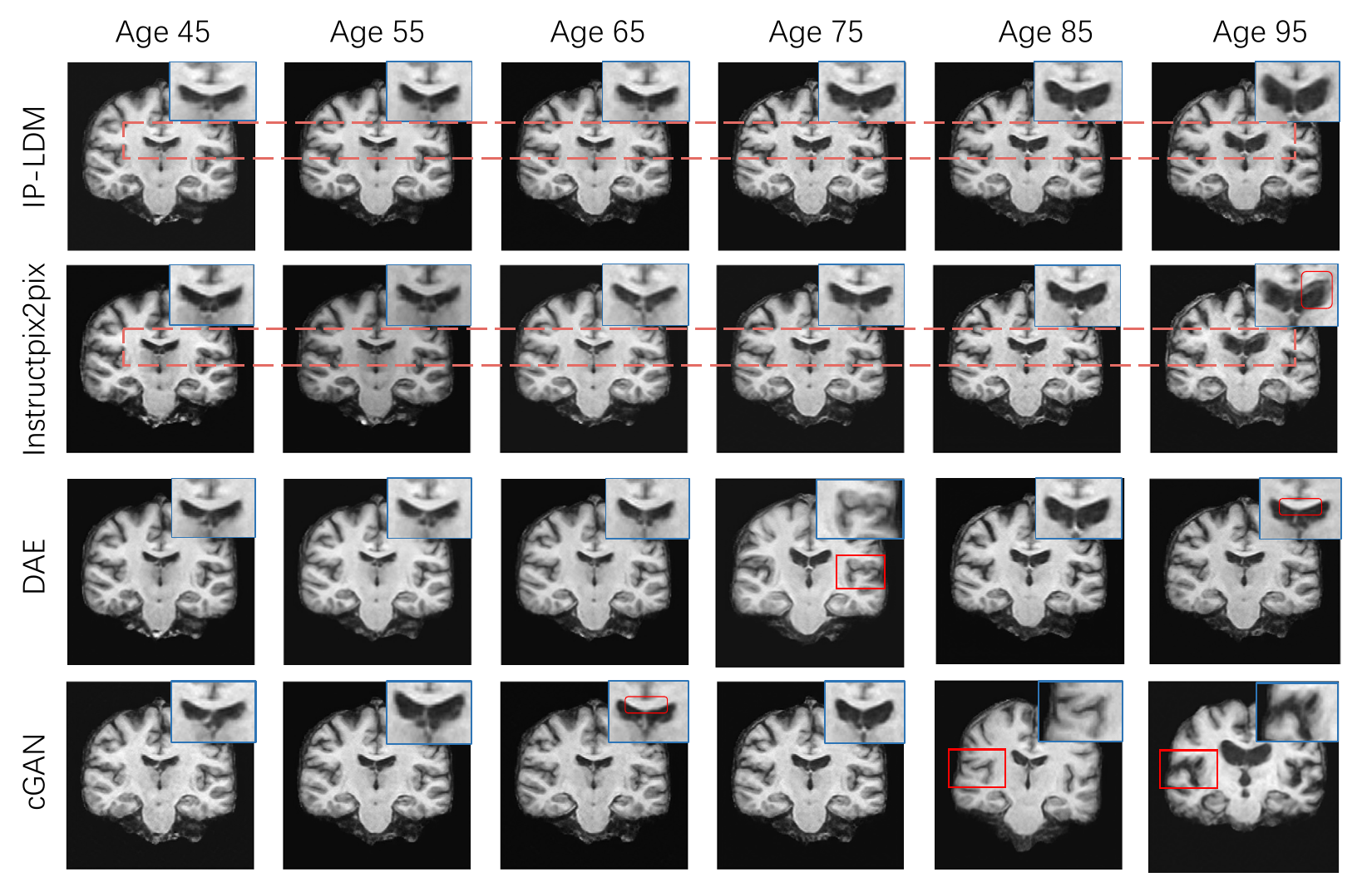}
		\vspace{-13pt}
		\caption{Qualitative visualization for brain age transformation. Each row represents brain images generated at different ages. The first row showcases the results from the proposed method IP-LDM. Subsequent rows display results from other methods, including InstructPix2Pix, DAE, and cGAN. Each column represents brain images generated at specific ages, ranging from age 35 to age 85.}
		\label{fig:q1}
	\end{figure*}

	Fig. ~\ref{fig:q1} illustrates the qualitative visualization results of brain image generation across varying ages on different methods.
	The first row showcases the results of IP-LDM, exhibiting a remarkable ability to maintain the structural integrity and unique features of the brain across all age transformations. Notably, IP-LDM successfully generates brain images with ventricles that grow larger with increasing age, while maintaining subtle geometric variations specific to each subject. This behavior is consistent with the transformation of brain aging observed in prior studies~\cite{peng2022gate, blinkouskaya2021brain}. These results underscore the model's effectiveness in reflecting the natural anatomical changes associated with brain aging while preserving individual identity. In contrast, the other methods display varying degrees of identity preservation. Although InstructPix2Pix is capable of generating age-progressed images, often fails to precisely maintain the fine details and displays significant artifacts, e.g., at the age of 95. DAE, while focusing on age transformation, exhibits wrong transformation trending and inconsistencies in preserving structural features, leading to noticeable deviations in brain aging progress, e.g., at the age of 75 and 95. Similar to DAE, Although the cGAN produces visually realistic images, it struggles with maintaining the precise age transformation, such as at the age of 65, and identity-specific features, such as at the age of 85 and 95.
	
	The quantitative analysis (as depicted in Tab.~\ref{table2}) further reinforces the superiority of IP-LDM by comparing the performance metrics (FID and KID) across different methods. The proposed IP-LDM model consistently achieves the lowest average scores for both FID (4.488) and KID ($0.623\times 10^{-5}$), indicating superior image quality and fidelity. Across various age ranges, IP-LDM maintains lower FID, particularly excelling in the 71-80 and 81-90 age groups with FID scores of 3.287 and 3.822, respectively. Similarly, IP-LDM achieves the lowest KID scores across all age ranges. These results, combined with the qualitative visualization, highlight the superiority and robustness of IP-LDM in preserving the identity of the brain during age transformation.

	\begin{table*}[ht]
\begin{center}
\scalebox{0.75}{
    \addtolength{\tabcolsep}{-0.6pt}
    \begin{tabular}{l|ccccc|ccccc} 
    \toprule
    \multirow{2}{*}{\textbf{Method}} & \multicolumn{5}{c|}{\textbf{OASIS-3}} & \multicolumn{5}{c}{\textbf{Baby Brain}}\\
    \cline{2-11}
     & SSIM $\uparrow$ & PSNR $\uparrow$ & FID $\downarrow$ & RMSE ($10^{-2}$)$\downarrow$ & ARI & SSIM $\uparrow$ & PSNR $\uparrow$ & FID $\downarrow$ & RMSE ($10^{-2}$)$\downarrow$ & ARI\\
    \midrule
    cGAN\cite{ren2021q} & 0.920 & 31.824 & 8.313 & 3.240 & 0.85 &  0.651 & 32.074 & 9.320 & 6.743 & 0.83  \\
    DAE\cite{preechakul2022diffusion}  & 0.912 & 27.08 & 7.296 & 2.361 & 0.93 & 
    0.616 & 29.560 & 9.765 &  5.514 & 0.81 \\
    InstructPix2Pix\cite{brooks2023instructpix2pix} & 0.940 & 34.35 & 5.972 & 2.037 & 0.99 & 
    0.381 & 30.832 & 6.714 & 12.68 & 0.96  \\
    \midrule
     IP-LDM   & 0.949 & 35.15 & 4.733 & 1.868 & 0.99 & 
     0.674 & 32.989 & 4.984 & 8.996 & 0.99 \\
    \bottomrule
    \end{tabular}}

\end{center}
    \caption{Comprehensive performance comparison of different methods on the OASIS-3 and IBIS datasets. The evaluation metrics used are Structural Similarity Index (SSIM), Peak Signal-to-Noise Ratio (PSNR), Fréchet Inception Distance (FID), Root Mean Square Error (RMSE), and Adjusted Rand Index (ARI).}
    \label{table1}
\vspace{-7pt}
\end{table*}

	\begin{table*}[h]
\begin{center}
\small
\scalebox{0.90}{
\addtolength{\tabcolsep}{1.5pt}
\begin{tabular}{l|c|ccccccc} 
\toprule
Metric & \diagbox{Methods}{Age} & 40-50 & 51-60 & 61-70 & 71-80& 81-90 & 91-100 & Avg.\\
\hline
\multirow{4}{*}{FID $\downarrow$ } 
& cGAN\cite{ren2021q} & 8.711 & 7.943 & 7.789 & 7.033 & 7.134 & 8.661 & 7.878 \\
& InstructPix2Pix\cite{brooks2023instructpix2pix} & 4.336 & 4.883 & 4.883 & 4.020 & 4.828 & 6.821 & 4.962 \\
    
    & DAE\cite{preechakul2022diffusion} & 6.441 & 5.318 & 5.661 & 5.192 & 5.377 & 6.073 & 5.677 \\
    \cline{2-9}
    & \textbf{IP-LDM} & 5.467 & 4.863 & 4.353 & 3.287 & 3.822 & 5.134 & 4.488 \\
\hline
\multirow{4}{*}{KID ($10^{-5}$)$\downarrow$ } & cGAN\cite{ren2021q} & 7.891 & 6.731 & 6.513 & 5.414 & 5.139 & 7.311 & 6.499 \\
& InstructPix2Pix\cite{brooks2023instructpix2pix} & 1.211 & 1.677 & 1.226 & 1.080 & 1.358 & 1.852 & 1.401 \\
    
    & DAE\cite{preechakul2022diffusion} & 1.734 & 1.661 & 1.419 & 1.319 & 1.355 & 1.891 & 1.563 \\
    \cline{2-9}
    & \textbf{IP-LDM} & 1.167 & 0.720 & 0.543 & 0.463 & 0.456 & 0.387  & 0.623 \\
    
\bottomrule
\end{tabular}}
\vspace{-4pt}
\end{center}
\caption{Quantitative analysis of age transformation comparison on the OASIS3 and Baby Brain datasets. Metrics evaluated include FID and KID across various age ranges (40-50, 51-60, 61-70, 71-80, 81-90, 91-100). IP-LDM consistently achieves the lowest average scores for both FID and KID, demonstrating superior performance in generating high-quality, realistic, and identity-preserving brain images.}
\label{table2}
\vspace{-7pt}
\end{table*}
	\begin{table*}[h]
\begin{center}
\scalebox{0.75}{
\addtolength{\tabcolsep}{0.1pt}
\begin{tabular}{lllll|cccc|cccc} 

\toprule
    \multirow{2}{*}{Config} & 
    \multirow{2}{*}{CC} & \multirow{2}{*}{CN} & \multirow{2}{*}{IL} & \multirow{2}{*}{IP} & \multicolumn{4}{c|}{\textbf{OASIS-3}} & \multicolumn{4}{c}{\textbf{Baby Brain}}\\
\cline{6-13}
& & & & & SSIM $\uparrow$ & PSNR $\uparrow$ & FID $\downarrow$ & RMSE ($10^{-2}$)$\downarrow$ & SSIM $\uparrow$ & PSNR $\uparrow$ & FID $\downarrow$ & RMSE ($10^{-2}$)$\downarrow$ \\
\midrule

A & \Checkmark & \XSolidBrush & \XSolidBrush & \XSolidBrush & 0.940 & 34.35 & 5.972 & 2.037 & 
0.551 & 31.94 & 5.842 & 11.68 \\
B & \Checkmark & \Checkmark & \XSolidBrush & \XSolidBrush & 0.944 &34.84 & 5.336 & 1.968 & 
0.596 & 32.49 & 5.629 & 11.32 \\
C & \Checkmark & \Checkmark& \Checkmark& \XSolidBrush& 0.948 & 35.10 & 5.047 & 1.878 & 
0.612 & 32.79 & 5.388 & 10.64 \\
D & \Checkmark & \Checkmark & \Checkmark & \Checkmark & 0.949 & 35.15 & 4.773 & 1.868 & 
0.674 & 32.99 & 4.984  & 9.00 \\

\bottomrule
\end{tabular}}
\end{center}
\vspace{-4pt}
\caption{Ablation study that evaluates the impact of four different configurations (A, B, C, and D) on the performance of IP-LDM across the OASIS-3 and Baby Brain datasets. Configuration A involves the concatenation (``CC'') of the source image into the reverse diffusion process, configuration B employs an identity control network (``CN'') to guide the denoising U-Net, configuration C incorporates an identity loss (``IL'') function, and configuration D utilizes an identity projector (''IP``) within the identity preservation module.}
\label{table:ablations}
\vspace{-7pt}
\end{table*}
	
	\subsection{Ablation Studies} 

	In our study, we conduct a comprehensive analysis through four ablation configurations, designated as A, B, C, and D, which are as follows: 
	\underline{Configuration A} involves the concatenation of the source latent features into the reverse process, investigating how the direct concatenation influences the model's ability to reconstruct accurate brain images.
	\underline{Configuration B} utilizes an identity control network to guide the denoising U-Net, ensuring the preservation of the subject's features during the generation process.
	\underline{Configuration C} employs an identity loss to constrain the identity preservation module, helping to maintain the subject's unique features.
	\underline{Configuration D} utilizes an identity projector within the identity preservation module, serving as an intermediary representation to maintain identity features more effectively.
	
	The ablation results in Tab.~\ref{table:ablations} show that D achieves superior performance in terms of identity preservation and image quality across both the OASIS-3 and Baby Brain datasets. For the OASIS-3 dataset, D records the highest SSIM (0.949) and PSNR (35.15), and the lowest FID (4.773) and RMSE (1.868). Similarly, on the Baby Brain dataset, D also outperforms other configurations, achieving the highest SSIM (0.674) and PSNR (32.99), and the lowest FID (4.984) and RMSE (9.00). These results highlight the effectiveness of the identity projector in ensuring robust identity preservation and aligning with observed brain aging changes. B and C also show notable improvements over baseline A. For the OASIS-3 dataset, B improved SSIM to 0.944 and reduced FID to 5.336, while C further enhanced SSIM to 0.948 and lowered FID to 5.047. On the Baby Brain dataset, B improved SSIM to 0.596 and reduced FID to 5.629, with C showing further improvement in SSIM to 0.612 and FID to 5.388. These findings demonstrate the importance of precise control and quantitative enforcement in maintaining identity features during age transformation.
	


	\section{Conclusion}
	In this work, we present an age- and identity-conditioned latent diffusion model, IP-LDM, for brain image transformation. Our approach allows for the modification of a single input brain image to reflect arbitrary and continuous age changes while maintaining the subject's identity. 
	By integrating triplet contrastive constraints, we ensure consistent intra-subject identity representation across transformations. 
	Our extensive evaluations on both elderly and infant brain datasets confirm the effectiveness of our method in synthesizing high-quality brain aging transformations. This demonstrates the potential of our model for applications in medical imaging and neuroscience, providing a powerful tool for studying brain aging processes.
	There are limitations in our current work that will be addressed in future research: (1) different datasets necessitate retraining the model, and (2) extending the approach to 3D data. 
	For further details, please refer to App.~\ref{supp_sec:limitations}.


	\medskip




	
	
	\bibliographystyle{plain}
	\bibliography{ref}

	\newpage
	\appendix
	
	\section{AE Training and Reconstruction Results}

	\begin{figure*}[h]
		\centering
		\includegraphics[width=\linewidth]{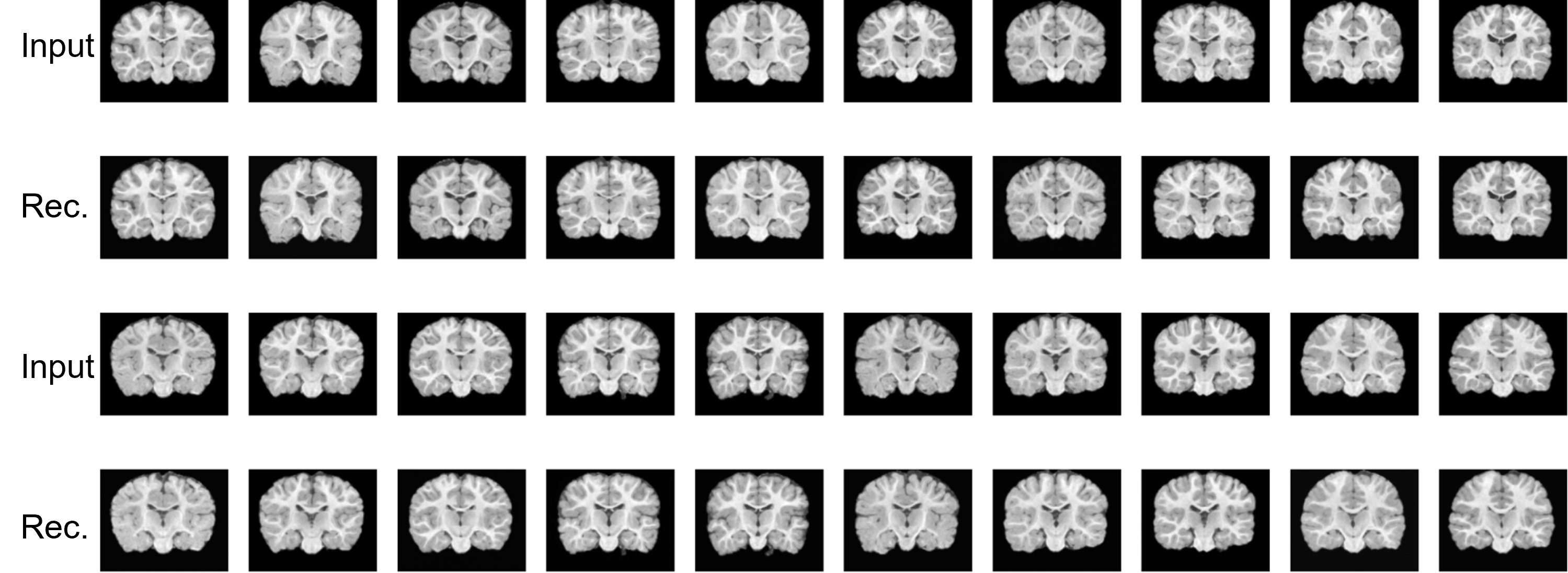}
		\caption{
			Inputs and reconstructions of VAE trained on Baby Brain.
			The first and third rows display the input baby brain MR images. 
			The second and fourth rows show the reconstructed baby brain images.
		}
		\label{baby_brain_vae_rec}
	\end{figure*}

	\begin{figure*}[h]
		\centering
		\includegraphics[width=\linewidth]{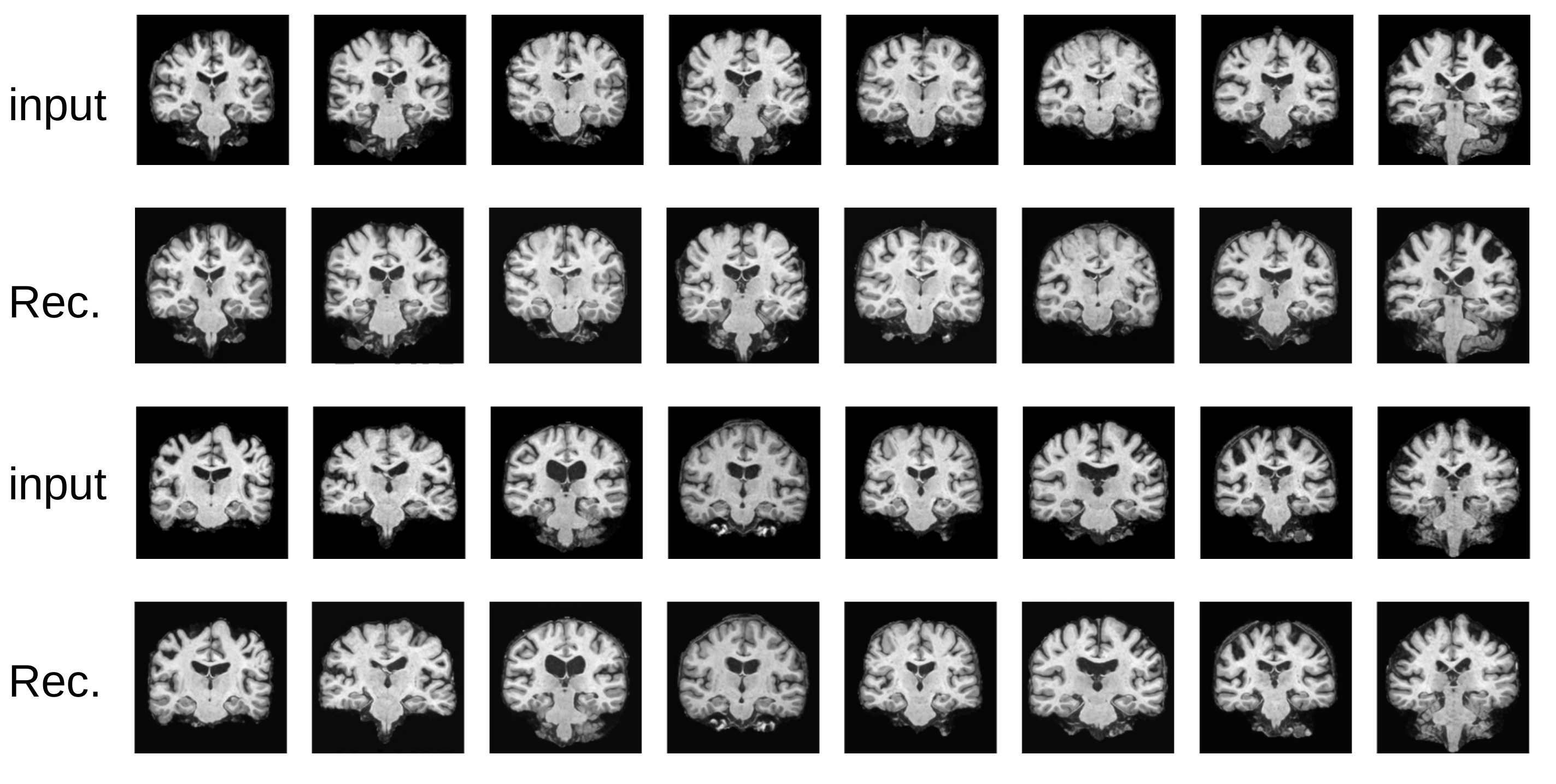}
		\caption{
			Inputs and reconstructions of VAE trained on OASIS-3.
			The first and third rows display the input OASIS-3 brain MR images. 
			The second and fourth rows show the reconstructed OASIS-3 brain images.
		}
		\label{oasis_vae_rec}
	\end{figure*}
	
	\begin{figure*}[h]
		\centering
		\includegraphics[width=\linewidth]{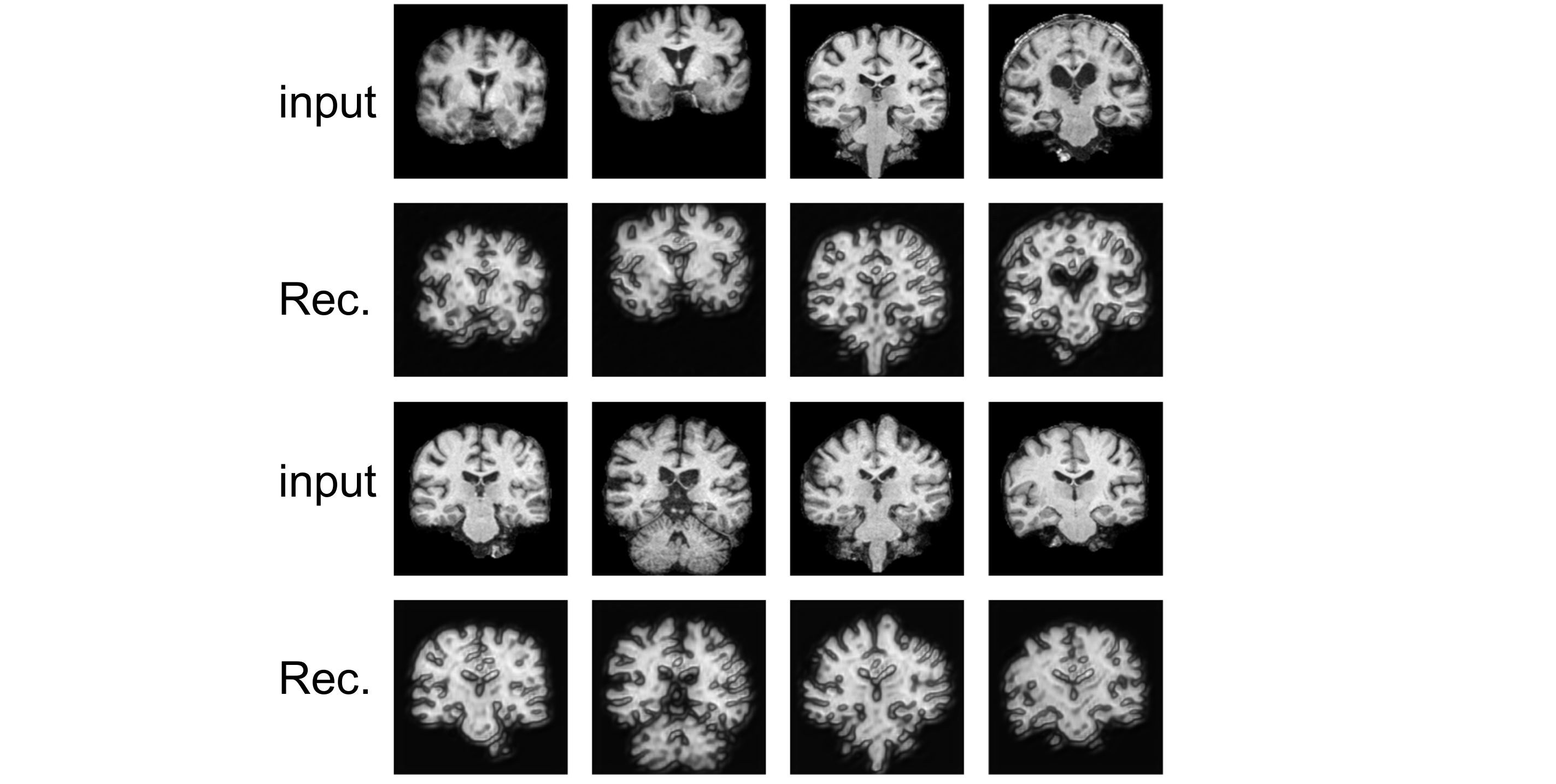}
		\caption{
			Effect of KL regularization on VAE reconstructions of OASIS-3 brain MR images.
			The first row displays the input brain MR images, while the second row shows the reconstructed images generated by the VAE. 
		}
		\label{oasis_vae_kl}
	\end{figure*}

	In Fig.~\ref{baby_brain_vae_rec} and Fig.~\ref{oasis_vae_rec}, we show the reconstruction results of our VAE model. 
	The results in the figure suggest that our VAE model can faithfully reconstruct the input images. 
	From the figure, we can observe that the VAE is able to reconstruct brain images at different developmental stages of infants. 
	This observation aligns with the known characteristic that infant brain images are usually less defined at younger ages and become sharper as the infant grows~\cite{almli2007nih}.

	\begin{figure*}[h]
		\centering
		\includegraphics[width=\linewidth]{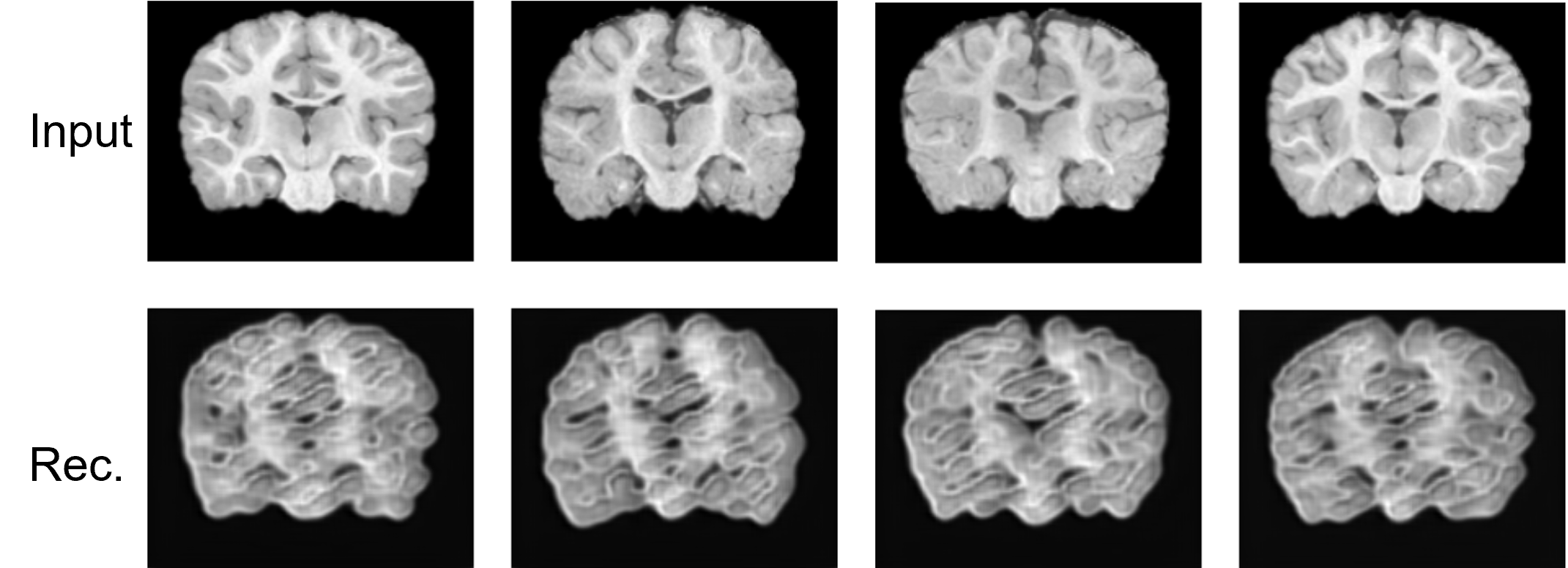}
		\caption{
			Effect of batch size on VAE reconstructions of Baby Brain MR images as discussed in Sec A. When the batch size reduces from 256 to 32, the reconstruction quality greatly degraded. 
			The first row displays the input baby brain MR images, while the second row shows the reconstructed images generated by the VAE. 
		}
		\label{baby_brain_vae_batch_size}
	\end{figure*}
	
	\textbf{AE KL Influence}
	In Fig.~\ref{oasis_vae_kl}, we demonstrate that training with a KL regularization significantly affects the performance of the VAE. With a KL regularization, the VAE generates less clear MR images, resulting in blurrier outputs. 
	
	\textbf{AE Batch Size Influence.}
	In Fig.~\ref{baby_brain_vae_batch_size}, we demonstrate that training with an inappropriate batch size significantly affects the performance of the VAE. 
	When the training batch size is reduced from 256 to 32, the VAE generates less clear MR images, resulting in blurrier outputs. 
	Additionally, the generated images exhibit pattern collapse, where the output patterns become overly similar.

	
	\section{More Longitudinal Generation Results}
	\begin{figure*}[h]
		\centering
		\includegraphics[width=\linewidth]{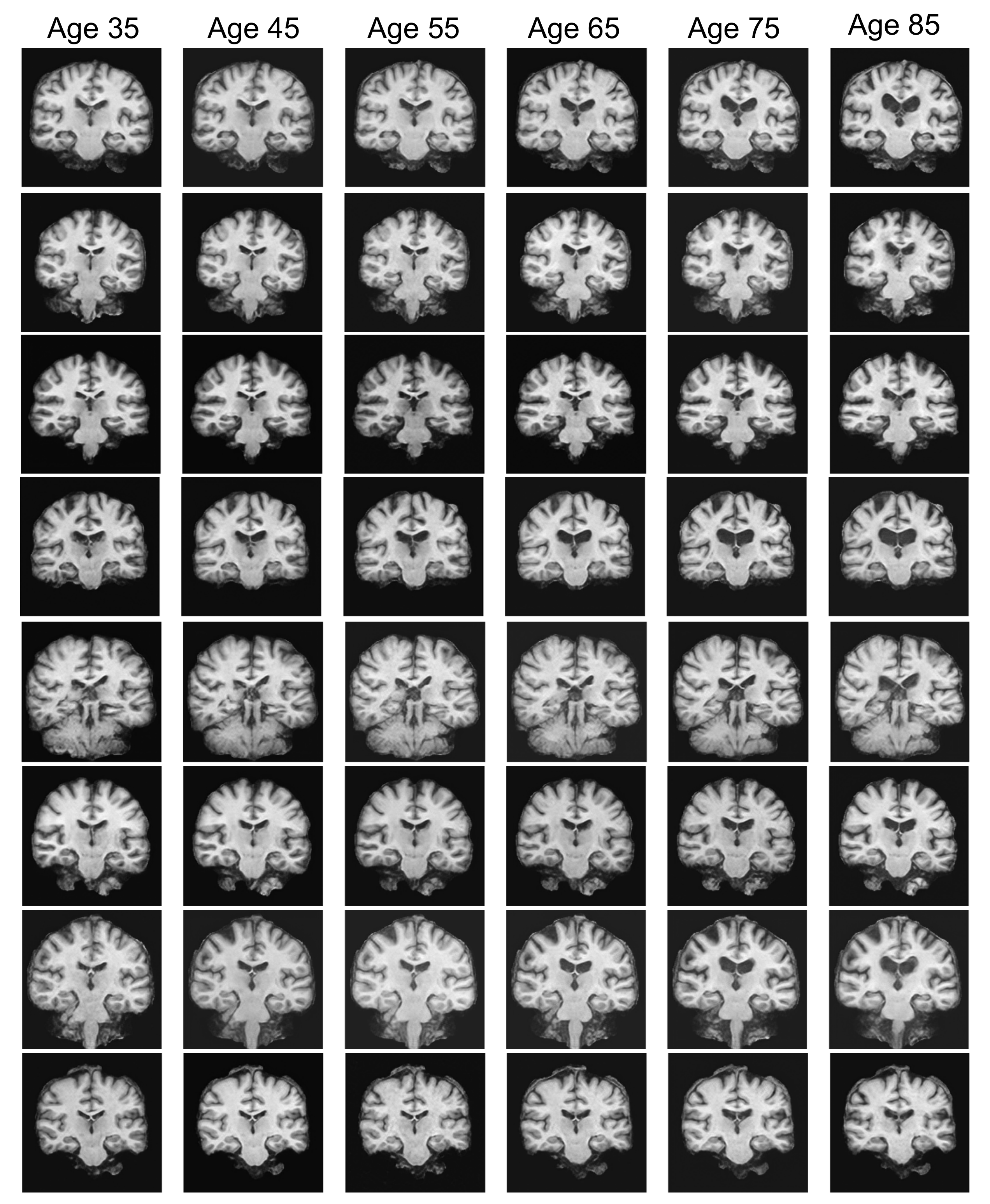}
		\caption{More longitudinal generation results on OASIS-3.
		}
		\label{oasis_more}
	\end{figure*}

	More results of longitudinal brain age transformation are illustrated in Fig.~\ref{oasis_more}. The figure clearly demonstrates IP-LDM's ability to generate realistic age-progressed images that maintain key structural features and individual-specific characteristics. Notably, the ventricles in the generated images grow larger with increasing age, While preserving the subtle geometric variations unique to each subject.

	\begin{figure*}[ht]
		\centering
		\includegraphics[width=\linewidth]{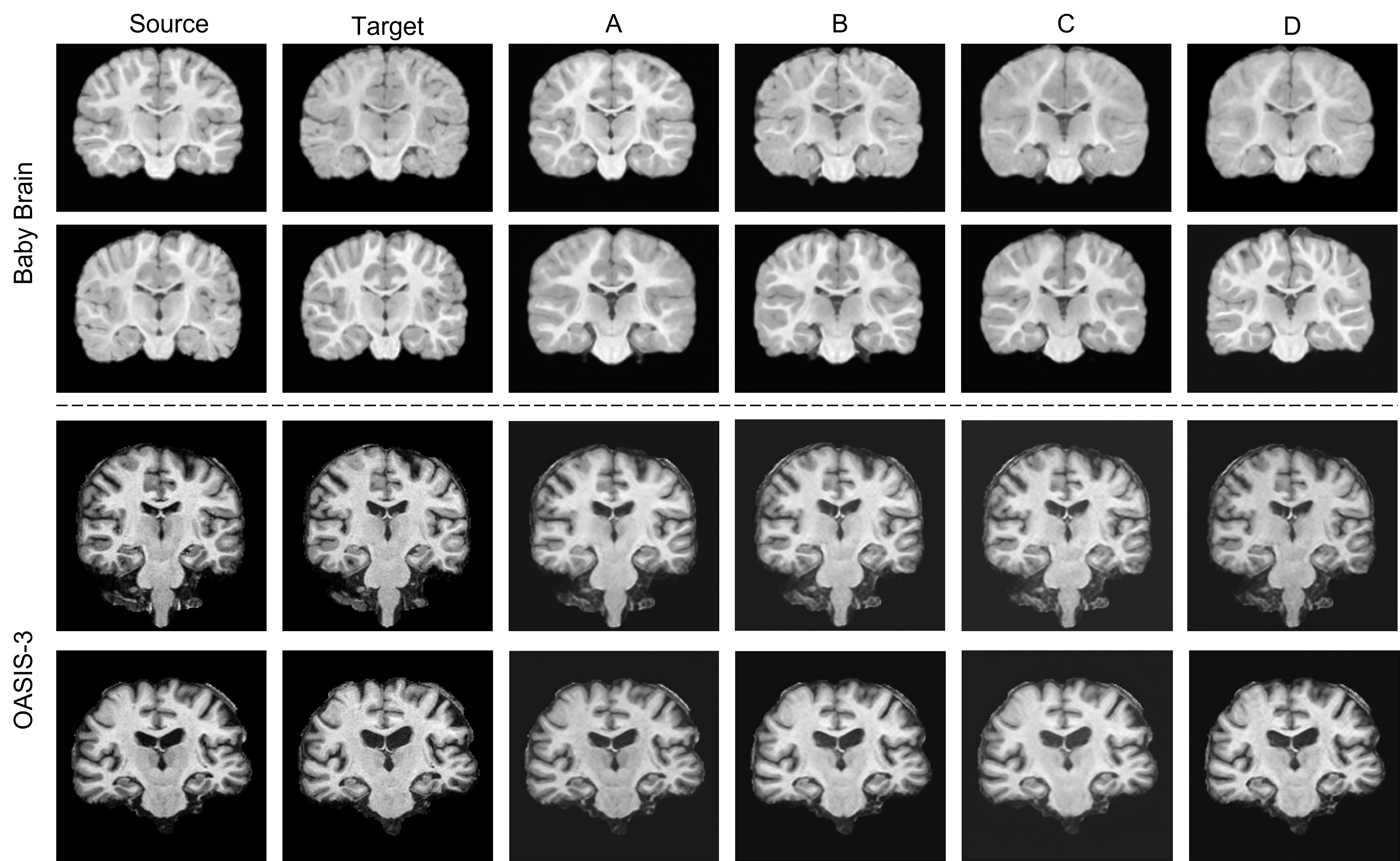}
		\caption{Visualization of the ablation study results on the OASIS-3 and Baby Brain datasets. Each row represents the brain images generated by different configurations (A, B, C, and D). }
		\label{fig:ablation}
	\end{figure*}
	Additionally, the visual results of the ablation study, shown in Fig.~\ref{fig:ablation}, further support the quantitative findings. For the OASIS-3 dataset, A shows moderate performance but struggles with finer identity details, leading to blurry artifacts.  B improves visual consistency and maintains identity features more effectively.  C further enhances identity preservation, while D exhibits the most superior performance, achieving accurate age transformation and preserving fine structural details and identity. Similar patterns are observed in the Baby Brain dataset, where D generates the most realistic and precise age transformations, effectively preserving subtle geometric variations and identity features.

	\section{Metrics}
	\label{supp_sec:metrics}
	We evaluate the generated images using the following metrics: Structural Similarity Index (SSIM), Peak Signal-to-Noise Ratio (PSNR), Fréchet Inception Distance (FID), Root Squared Mean Square Error (RSMSE), and Adjusted Rand Index (ARI).
	
	1. \textbf{Structural Similarity Index (SSIM)}:
	\[
	\text{SSIM}(x, y) = \frac{(2\mu_x\mu_y + C_1)(2\sigma_{xy} + C_2)}{(\mu_x^2 + \mu_y^2 + C_1)(\sigma_x^2 + \sigma_y^2 + C_2)}
	\]
	where \(\mu_x\) and \(\mu_y\) are the average of \(x\) and \(y\), \(\sigma_x^2\) and \(\sigma_y^2\) are the variance of \(x\) and \(y\), \(\sigma_{xy}\) is the covariance of \(x\) and \(y\), and \(C_1\) and \(C_2\) are constants to stabilize the division.
	
	2. \textbf{Peak Signal-to-Noise Ratio (PSNR)}:
	\[
	\text{PSNR} = 10 \log_{10} \left( \frac{\text{MAX}_I^2}{\text{MSE}} \right)
	\]
	where \(\text{MAX}_I\) is the maximum possible pixel value of the image and \(\text{MSE}\) is the mean squared error between the original and generated images.
	
	3. \textbf{Fréchet Inception Distance (FID)}:
	\[
	\text{FID} = \left\| \mu_r - \mu_g \right\|^2 + \text{Tr}(\Sigma_r + \Sigma_g - 2\sqrt{\Sigma_r \Sigma_g})
	\]
	where \((\mu_r, \Sigma_r)\) and \((\mu_g, \Sigma_g)\) are the mean and covariance of the real and generated image feature vectors, respectively.
	
	4. \textbf{Kernel Inception Distance (KID)}:
	\[
	\text{KID} = \frac{1}{n(n-1)} \sum_{i \neq j} k(\mathbf{x}_i, \mathbf{x}_j) + \frac{1}{m(m-1)} \sum_{i \neq j} k(\mathbf{y}_i, \mathbf{y}_j) - \frac{2}{nm} \sum_{i, j} k(\mathbf{x}_i, \mathbf{y}_j)
	\]
	Here, $k(\mathbf{x}, \mathbf{y}) = ( \mathbf{x}^\top \mathbf{y} + c )^d$ is the polynomial kernel function, where $c$ is a constant and $d$ is the degree of the polynomial.
	
	5. \textbf{Root Mean Square Error (RMSE)}:
	\[
	\text{RMSE} = \sqrt{\frac{1}{N} \sum_{i=1}^{N} (x_i - y_i)^2}
	\]
	where \(N\) is the number of pixels, and \(x_i\) and \(y_i\) are the pixel values of the original and generated images, respectively.
	
	6. \textbf{Adjusted Rand Index (ARI)}:
	\[
	\text{ARI} = \frac{\text{RI} - \text{Expected RI}}{\text{Max RI} - \text{Expected RI}}
	\]
	where \(\text{RI}\) is the Rand Index, and the expected and maximum values are adjusted for chance.

	\section{Dataset Age Distributions}
	\label{supp_sec:age_dist}

	\begin{figure*}[h]
		\centering
		\includegraphics[width=\linewidth]{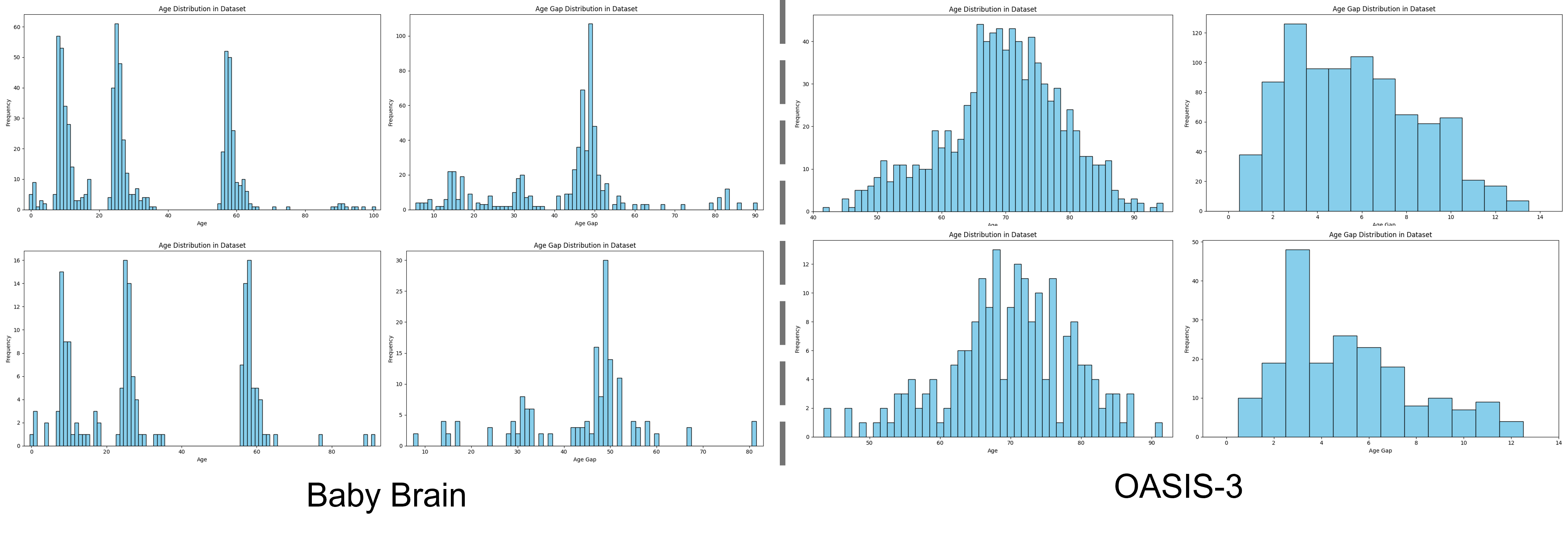}
		
		\caption{
			Age and age gap distributions of Baby Brain and OASIS-3 datasets.
			The top panels represent the training data, and the bottom panels represent the validation data for each dataset. The left panels show the age and age gap distributions for the Baby Brain dataset, highlighting three main age groups (6, 12, and 24 months) and age differences clustered around 12 months. 
			For Baby Brain, the x-axis is normalized from 0 to 36 months to 0 to 100. The right panels display the distributions for the OASIS-3 dataset.
		}
		\label{age_dist}
	\end{figure*}
	
	In Fig.~\ref{age_dist}, we show the age distribution of our two datasets, OASIS-3 and Baby Brain. From this figure, we can observe two key aspects of these datasets. First, the overall age distribution: for Baby Brain, there are three main regions corresponding to 6 months, 12 months, and 24 months. 
	In contrast, the age distribution for OASIS-3 resembles a normal distribution centered around 70 years, with a range from 40 to 95 years. 
	Second, the age gap distribution within specific individuals: Baby Brain's age differences are clustered around 12 months, while OASIS-3 exhibits a uniform distribution of age differences ranging from 1 to 10 years. 
	These differences in dataset characteristics affect the performance of diffusion-based and GAN-based baselines. 
	Despite these variations, our model consistently outperforms the baselines, demonstrating the robustness of our designed modules. 
	Additionally, because we randomly split the train and validation datasets, the distributions of these datasets follow the same trends.
	
	\section{Limitations}
	\label{supp_sec:limitations}
	Our dataset size is still relatively small compared to larger models, which limits the robustness and generalizability of our method. 
	This limitation highlights the need for gathering more data and pretraining the model in future work.
	Additionally, our method has the potential to be applied to 3D images, but this has not yet been explored. 
	This limitation will need to be addressed in future research.
	These two datasets are very similar in modality. However, in our experiments, merging them into one dataset did not result in better convergence. The evidence suggests that merging the datasets actually hindered the convergence process.
	
	\section{Broader Impact}
	This paper presents work aimed at advancing the field of Machine Learning. We acknowledge that our work may have various societal consequences, although we do not feel the need to highlight any specific ones here.

\end{document}